\documentclass[a4paper]{article}
\usepackage{mystyle}
\usepackage{float}

\title{Differences in boundary behavior in the 3D vertex and Voronoi models}
\author[1]{E. Lawson-Keister}
\author[2]{Tao Zhang}
\author[1]{M. L Manning}
\affil[1]{Department of Physics and BioInspired Syracuse, Syracuse University, Syracuse, NY 13244, USA}
\affil[2]{School of Chemistry and Chemical Engineering, Shanghai Jiao Tong University,
Shanghai 200240, China}

\begin{document}
\maketitle

\begin{abstract}
An important open question in the modeling of biological tissues is how to identify the right scale for coarse-graining, or equivalently, the right number of degrees of freedom. For confluent biological tissues, both vertex and Voronoi models, which differ only in their representation of the degrees of freedom, have effectively been used to predict behavior, including fluid-solid transitions and cell tissue compartmentalization, which are important for biological function. However, recent work in 2D has hinted that there may be differences between the two models in systems with heterotypic interfaces between two tissue types, and there is a burgeoning interest in 3D tissue models. Therefore, we compare the geometric structure and dynamic sorting behavior in mixtures of two cell types in both 3D vertex and Voronoi models. We find that while the cell shape indices exhibit similar trends in both models, the registration between cell centers and cell orientation at the boundary are significantly different between the two models. We demonstrate that these macroscopic differences are caused by changes to the cusp-like restoring forces introduced by the different representations of the degrees of freedom at the boundary, and that the Voronoi model is more strongly constrained by forces that are an artifact of the way the degrees of freedom are represented. This suggests that vertex models may be more appropriate for 3D simulations of tissues with heterotypic contacts.
\end{abstract}

\section{Introduction}

Mechanical interactions between molecules, cells, and tissues are increasingly being identified as control mechanisms in development and disease.  In order to make quantitative predictions about how physical forces impact processes such as tissue compartmentalization and cell migration in dense cellularized tissues, it is necessary to develop well-vetted mechanical models. In this work, we focus on models for confluent tissues, such as epithelial layers, where there are no gaps or overlaps between cells.

Traditionally, much of the computational work to simulate confluent tissue has been restricted to two-dimensional (2D) models~\cite{Bi2015, Bi2016, Sahu2020, Sahu2021, Graner1992, Farhadifar2009, Staple2010, Lawson-Keister2021}. While 2D models for confluent tissue are extremely powerful, they rely on several assumptions in order to approximate collections of cells that exist in three-dimensional (3D) space. One assumption is that all the cells of interest exist in a single monolayer, and that the mechanics of that monolayer is dominated by interactions in a single (usually apical) 2D plane. Another assumption is that any interactions with the environment, such as a basement membrane, can be approximated by body forces acting on cell centers or vertices.  Finally, it is assumed that any fluctuations or dynamics in the unresolved third dimension (such as fluctuations in height) can be mapped in a simple way onto dynamics in the two-dimensional plane (such as fluctuations in cross-sectional area). But clearly, these assumptions are not always valid, and there are many systems that require a fully three-dimensional model such as in organ/organoid development~\cite{Sanematsu2021, Zhang2022}, cancer spheroids~\cite{Zanoni2016, Nath2016, Paradiso2021}, and cell sorting in 3D cellular aggregates~\cite{Hutson2008, Sahu2021}. 

In addition, compartmentalization of different tissue types and formation of boundaries between different cell types is crucial to proper function in development~\cite{Rubsam2018, Fuchs2007, Blanpain2009, Tinkle2008, Vasioukhin2001, Unbekandt2010, Waites2018, Montero2005}, and treatment of disease~\cite{Song2016, Friedl2009}. While the exact mechanisms that drive compartmentalization and boundary formation are still an area of active research, several simple hypotheses have helped to drive forward the field over the past 60 years. The differential adhesion hypothesis (DAH)~\cite{M.S.Steinberg1963, Foty2005} postulates that cell sorting is driven by differences in cell surface tensions, which in turn arise from differences in cell adhesion. This hypothesis correctly explains behavior in particle-based simulations of cells where particles interact with adhesion-like terms that depend on the distance between cell centers~\cite{Buhn2006, Safran2002}.

In experimental observations of confluent tissues, where cells are not well-represented as sticky spheres, cell-cell adhesion must instead influence cell shapes and cell interfacial tensions. The differential interfacial tension hypothesis (DITH)~\cite{Harris1976} hypothesizes that differences in interfacial tension between two cell types drive cell sorting, and suggests that adhesion and surface contractility compete to create an effective interfacial tension between cells of opposite types, sometimes called a heterotypic interfacial tension~\cite{Canty2017, Sussman2018a, Barton2017, Graner1992}. Importantly, DITH and DAH are not irreconcilable, as in many tissue types there are signaling feedbacks between adhesion and cortical tension~\cite{Yamada2007} that lead to a down-regulation of cortical tension at interfaces with increased adhesion, so that both DITH and DAH predict similar outcomes~\cite{Amack2013, Maitre2012b}.

In confluent tissues, adhesion expression governs both individual cell shapes in a monolayer~\cite{Sahu2020} \emph{and} heterotypic interactions between two different cell types~\cite{Canty2017,Tsai2020}, and it is not immediately obvious whether differences in cell shape alone are also sufficient to generate sorting.  Recent experiments and simulations have suggested that at least in cell monolayers, differences in cell shape can generate small-scale segregation but not large-scale demiximing~\cite{Sahu2020}.

In contrast, simulations with heterotypic interfacial tension exhibit robust large-scale cell sorting~\cite{Graner1992} with sharp boundaries between tissue types, which closely mirrors the dynamics and boundaries observed in experimental tissues~\cite{Landsberg2009, Dahmann2011, Nnetu2012, Calzolari2014}. It is difficult to quantify heterotypic interfacial tensions directly \emph{in situ}, although laser ablation approaches~\cite{Sugimura2016} can approximate differences in tension up to an assumed viscous constant. Alternatively, pipette aspiration experiments~\cite{Rubsam2017, Maitre2012b, Tsai2020a} can study tension properties of extracted cell doublets or triplets.

Recent simulation work has focused on understanding how the geometry of heterotypic interfaces impacts their mechanical properties. A study focused on 2D Voronoi and vertex models demonstrated that several different definitions of surface tension that are all equivalent in equilibrium fluids are not the same in confluent tissue models. Heterotypic surface tension along interfaces suppresses capillary waves, resulting in sharp yet deformable boundaries~\cite{Sussman2018a}, with characteristic geometric signatures in cells near the interface~\cite{Sahu2020}. Interestingly, both effects appear to be less strong in vertex compared to Voronoi models in 2D. In 3D Voronoi model simulations, similar geometric signatures are observed including elongation of cells on the boundary, sharp interfaces, and compartmentalization of cells of opposite types~\cite{Sahu2021}. 

Interestingly, a recent paper observes a small difference in the geometric behavior of cells on a heterotypic boundary in the 2D vertex and Voronoi models~\cite{Sahu2021}. In the Voronoi model, cells across a heterotypic interface will align their cell centers, while in the 2D vertex model, they align significantly less. This presents a challenge to Voronoi models, as the choice of model degrees of freedom should not change predictions about tissue morphology if both models use the same energy functional and are meant to encapsulate the same physics.

Quite a lot of 3D modeling of confluent tissues has been performed using 3D Voronoi models~\cite{Merkel2018, Sanematsu2021, Li2018, Grosser2021}, mainly due to their computational efficiency and simplicity compared to 3D Vertex models. Almost two decades ago, a first 3D vertex model was described and simulated~\cite{Honda2004}, and more recent work has explored other aspects of tissue mechanics with 3D vertex models~\cite{Okuda2013, Okuda2021, Hannezo2014}, though the code was not made available open source, and there are significant subtleties and challenges in handling changes to topology in 3D vertex models, whereas those are handled automatically in Voronoi code bases. Very recently, Zhang and Schwarz have reported on rigidity transitions and topological protection in 3D vertex models for organoids, and concurrently released an open-source 3D vertex model code~\cite{Zhang2022}, which permits a clear view of the coding choices made and is a significant contribution to the field.

Given the small difference seen between cell geometries in 2D vertex and Voronoi models, and the important role cell geometries play in both the physics of interfacial tension and in comparing model predictions to experimental observations, the obvious open question is whether there are any geometric differences near heterotypic interfaces in 3D vertex and Voronoi models. Therefore, we adapt the 3D vertex simulation code developed by~\cite{Zhang2022} to investigate cell shape and dynamics at heterotypic interfaces and compare that to the 3D Voronoi model to discern whether there are differences and, if so, explain the mechanisms that drive those differences.

\section{Model and Methods}
For both 3D vertex and Voronoi models, the simulated confluent tissue is composed of $N = 1728$ cells driven by active forces. Each cell $i$ is represented by polyhedra with mechanics that are driven by the energy functional 
\begin{equation}
E= \sum_i^N K_{A}(A_i-A_{0})^2+\sum_i^N K_{V}(V_i-V_{0})^2  + E_0,
\label{eqn:EnergyBase3D}
\end{equation}

where the mechanical interaction forces between cells are given by $F_i=-\nabla_i E$. The first term describes the competition between surface tension and adhesion where cell-cell contacts are represented by intercellular areas and have area modulus $K_A$. The second term represents the cells trying and achieve a characteristic cell size of volume $V_0$ with volumetric modulus $K_{V}$. Here we are explicitly relaxing the constraint that cells are incompressible and suggesting that a combination of biological mechanisms, such as ion channels, regulate the volume to stay within a range parameterized by $K_{V}$. Then both systems evolve under Brownian dynamics. The only difference between the two models is the degrees of freedom in the Voronoi model are the cell centers and the degrees of freedom in the vertex model are the cell vertices.

A three-dimensional non-dimensionalized shape index can be defined as $S_0 = A_0 / V_0^{2/3}$. As the shape index of a cell increases cells will become less circular or more elongated. Both models experience a rigidity transition in homogenous systems in which the tissue goes from behaving solid-like to fluid-like as a function of the cell shape. In the 3D Voronoi model, this rigidity transition occurs at a cell shape of $s_0 = 5.413$ and is identified by the point in which the shear modulus vanishes~\cite{Merkel2017}. In the 3D Vertex model, the rigidity transition is identified by looking at the neighbor-overlap function and the tissue becomes fluid-like when the typical time-scale for rearrangements becomes zero which occurs at $s_0 = 5.39 \pm 0.01$~\cite{Zhang2022}. A similar agreement in the location of rigidity transition with cell shape is seen in the 2D vertex and Voronoi models~\cite{Bi2015, Bi2016}.

While homogeneous tissue seems to behave quite similarly between the two models, it is also important to investigate tissue consisting of multiple cell types. As previously stated, the addition of an additional energy cost for heterotypic interfaces can cause large-scale demixing and compartmentalization between different cell types~\cite{Sahu2020}. This additional interfacial tension also induces a large difference in cell morphologies for cells at the interface between the tissue types. 

In the 2D vertex and Voronoi models the addition of interfacial tension between cells of different types changes the energy to
\begin{equation}
E= \sum_i^N K_{P}(P_i-P_{0})^2+\sum_i^N K_{A}(A_i-A_{0})^2  + E_0 + \sum_{HIT} \gamma_{ij} l_{ij},
\label{EnergyHITBase}
\end{equation}
where the sum is over all heterotypic interfaces, $l_{ij}$ is the interface between cell $i$ and $j$, and $\gamma_{ij}$ is the additional interfacial tension between cell types of cell $i$ and $j$. While a seemingly small term, it turns out that this provides a remarkably strong collective effect. This term causes cells of different types to quickly and robustly completely demix~\cite{Canty2017,Sussman2018a,Sahu2020} and can create sharp boundaries between cell types~\cite{Sussman2018a}. In completely confluent tissues these sharp boundaries generate topological consequences.

First, for a vertex on a completely flat boundary to be stable, it must have four neighbors instead of the typical three-fold coordination.   Since a stable vertex is under force balance and a flat interface will have two parallel edges, the third edge can never achieve force balance alone. This line of reasoning can be extended to prove that in homogeneous vertex models fourfold vertices are unstable~\cite{Su2016}, although in Voronoi models stabilized fourfold coordinated vertices have been observed at very high shape values. In contrast, the addition of heterotypic interfacial tension can regularly stabilize four-fold coordinated interfaces in both vertex and Voronoi models at all shape values~\cite{Sussman2018a}.

To investigate the configurations of cells on the boundary researchers quantified the distribution of edge lengths of cells on heterotypic interfaces in simulations of the 2D Voronoi model~\cite{Sussman2018a}. They find that there is an increasingly bimodal distribution of edge lengths as interfacial tension increases. This distribution is caused by perturbations to the stable interface due to finite temperature fluctuations. These perturbations will cause the predicted fourfold coordinated vertices to split and the single long-edge heterotypic interface will split into one large edge and one or two small edges.

In 2D Voronoi models, the breaking of these fourfold vertices generates a discontinuous restoring force that suppresses fluctuations. To quantify these discontinuities the researchers examined the restoring force generated by perturbing cells along a vector perpendicular to the interface. The authors then analytically calculated the restoring force on a Voronoi cell that is perturbed from a 9-cell square lattice. They found that the average restoring force cells experienced in simulations were very similar to that predicted by the analytic calculation and that the discontinuous restoring force scaled with the magnitude of the interfacial tension. Fig~\ref{fig:VerVorRestoring} illustrates some of these results for direct comparison to 2D vertex models.

The inclusion of heterotypic interfacial tension in 3D models is similar to that of 2D except the additional edge cost is for the shared surface area between cells of different types rather than edges.
\begin{equation}
E= \sum_i^N K_{A}(A_i-A_{0})^2+\sum_i^N K_{V}(V_i-V_{0})^2 + E_0  + \sum_{HIT} \sigma_{ij} A_{ij},
\label{eqn:EnergyHIT3D}
\end{equation}
where the sum is over interfaces between cells of different types, $\sigma_{ij}$ is the interfacial tension between the type of cell $i$ and the type of cell $j$ and $A_{ij}$ is the interfacial area between cell $i$ and cell $j$.

The authors of Ref~\cite{Sahu2021} investigate the geometric and dynamic signatures that arise from this additional interfacial tension. There is rapid demixing between cells of different types which causes the tissue to compartmentalize. Both the speed and magnitude of this demixing are determined by the magnitude of the interfacial tension. In addition, the cell shapes on the interface will start to elongate and orient perpendicular to the interfacial axis. Additionally, as cell shapes on the boundary increase as interfacial tension is increased, cells in the bulk will round up and decrease their cell shape. 

Cell orientation is calculated using the moment of inertia tensor of the best-fit ellipsoid to the cell vertices. Then the orientation is defined as the angle the long axis of the ellipse makes with the interface. The authors find that as heterotypic interfacial tension increases the cells go from random orientation as in the case with no HIT to highly oriented perpendicular to the axis of tension at high values of interfacial tension.

They also investigated a similar effect to what was seen in 2D which was the effect HIT had on the interfacial area along the interface. They observe a similar behavior to the 2D models, as the magnitude of the interfacial tension increase there is an increasingly bimodal distribution of small area facets and large area facets suggesting a similar phenomenon to the breaking of fourfold vertices in 2D.

Finally, the authors noticed that cells on the boundary start to resemble one-sided prisms with flat interfaces at the boundary. In a Voronoi model to achieve this geometry, cells on one side of the interface would need to align their centers in a plane parallel to the interface. Additionally, cells across the boundary must align their cell centers to minimize the distance between their centers in the plane parallel to the interface such that the cell centers become stacked or registered. This registration effect is defined in a system in which the heterotypic interface is in the XY plane in the following equation.
\begin{equation}
    R = 1 - \frac{d}{l_0}
\end{equation}
where $d$ is the distance between neighbors across the interface in the XY plane and $l_0$ is the average lattice spacing. If a cell is perfectly registered directly on top of its neighbor the registration will be unity and if they are perfectly misaligned half a lattice spacing away the registration will be zero. Sahu et al find that as the interfacial tension magnitude increases, the height of cells on the same side of the boundary converges and that registration goes from roughly uniformly distributed to being highly peaked near unity. 

In the paper supplement, they investigate this registration effect in the 2D models. They find that while the registration for the 2D Voronoi model shares similar behavior to the 3D Voronoi model, the 2D vertex model exhibits differences. The vertex model goes from a uniform registration to having a registration peak around a value of $R \approx 0.55$, which is distinct from uniform but obviously less than the value near unity seen in Voronoi models. The authors hypothesize that the difference is due to extra degrees of freedom in the vertex model which allows a relaxation of some of the constraints at the interface. But this poses the question: are the geometric signatures seen in the 3D Voronoi model robust to the choice of model?

\section{Results}
\subsection{Comparing vertex and Voronoi model structures in 3D}
To investigate the differences between these two models in three dimensions we adapt the 3D Voronoi model code used in Ref~\cite{Merkel2018} and the open-source 3D vertex model first published in Ref~\cite{Zhang2022}. First, we investigated the behavior of a 3D vertex model simulation comprised of two different cell types with heterotypic interfacial tension between them. Just as seen in 3D Voronoi models~\cite{Sahu2021}, cells rapidly segregate and become completely demixed, see Fig~\ref{fig:VertexDemixing}. The speed and magnitude of this demixing are increased as the magnitude of the interfacial tension is increased. Additionally, if the tissue starts in a completely demixed state the boundaries will remain stable and the tissue will stay demixed suggesting that this state is energetically preferred. 

\begin{figure}[ht!]
\includegraphics[width = 0.99\linewidth]{Figures/VertexDemixing.pdf}
\caption{
(A,B) Schematic of 2D Voronoi model (green, degrees of freedom are cell center) and 2D vertex model (purple, degrees of freedom are cell vertices). (C,D) Snapshot of the 3D Voronoi model (green) and vertex (purple) simulations with a heterotypic interface between light and dark cells. (E) Demixing behavior of the 3D Vertex model increasing HIT as the color becomes darker. (F) Demixing behavior between 3D Vertex and Voronoi models where the lighter colors start randomly initialized and the dark colors start sorted.}
\label{fig:VertexDemixing}
\end{figure}

Next, cell shapes in the tissue were examined. In the 3D Vertex model, the cells in the bulk of the tissue decrease their observed cell shape index as the magnitude of the heterotypic surface tension increases, consistent with the behavior observed in the Voronoi model. Similarly, cells on the boundary experience increases in cell shape with heterotypic tension, and the magnitude of the increase is even larger than the Voronoi model at large values of interfacial tension. This is likely due to the extra degrees of freedom in the vertex model allowing cells access to a wider range of cell shapes. 

\begin{figure}[ht!]
\includegraphics[width = 1.0\linewidth]{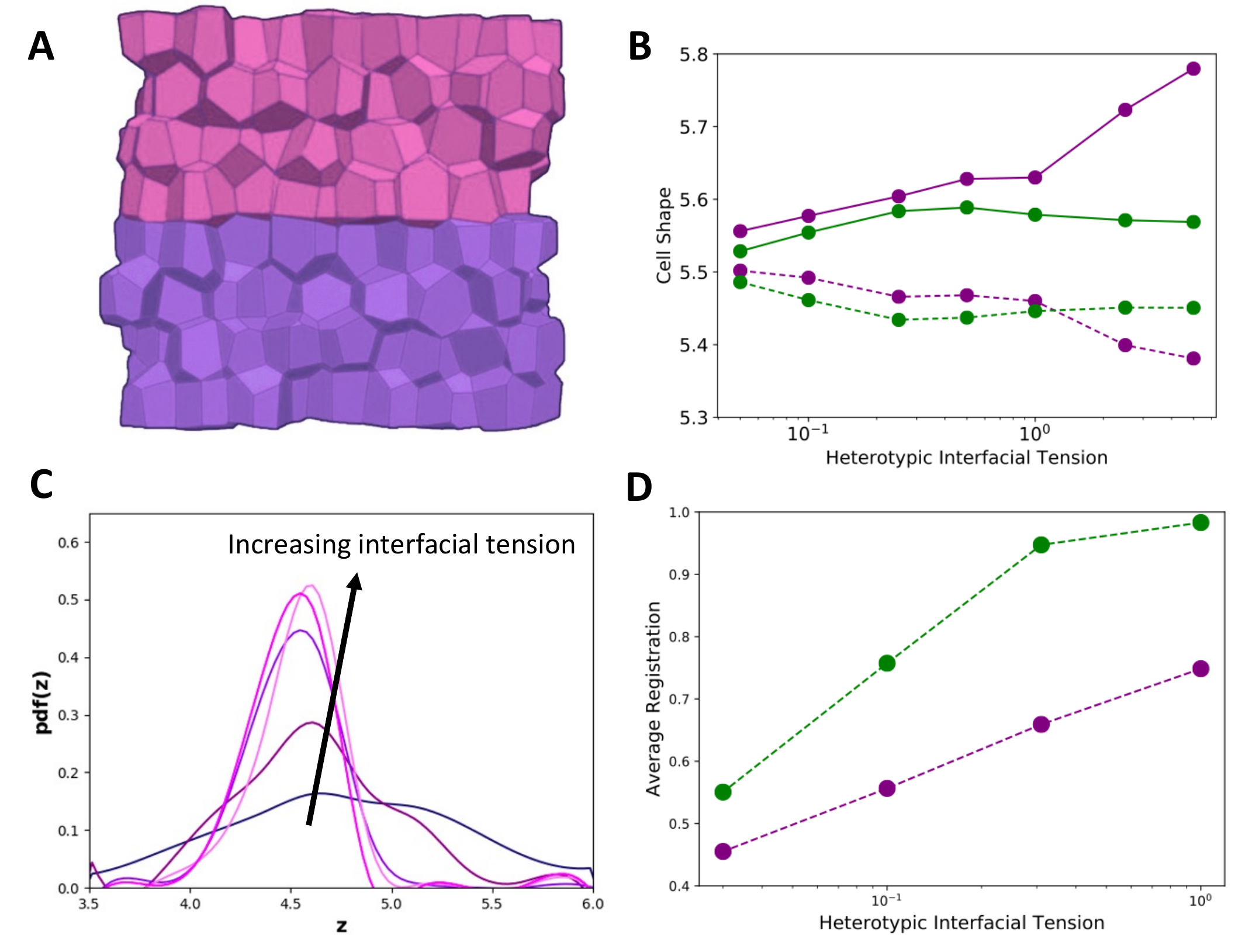}
\caption{
A.) Snapshot of the 3D Vertex model with HIT between cells of different types B.) Cells on the boundary of both models (vertex purple, Voronoi green), solid lines, increase cell shape while cells in the bulk, dashed lines, round-up with increasing HIT C.) Vertex cells on one side of the interface begin to align in at the same height as interfacial tension increases. D.) Registration increases for both models as interfacial tension increases.}
\label{fig:VertexGeometry}
\end{figure}

The cells along the boundary also exhibit similar registration behavior in both models. In the vertex model, cells on the same side of the interface will start to align their height in a plane as interfacial tension increases. These cells will also start to register with cells across the interface with an increasing magnitude as HIT increases. The magnitude of registration is significantly higher in the Voronoi model than in the vertex model, shown in Fig~\ref{fig:VertexGeometry}, and consistent with observations in the 2D models.

\begin{figure}[ht!]
\centering
\includegraphics[width = 0.9\linewidth]{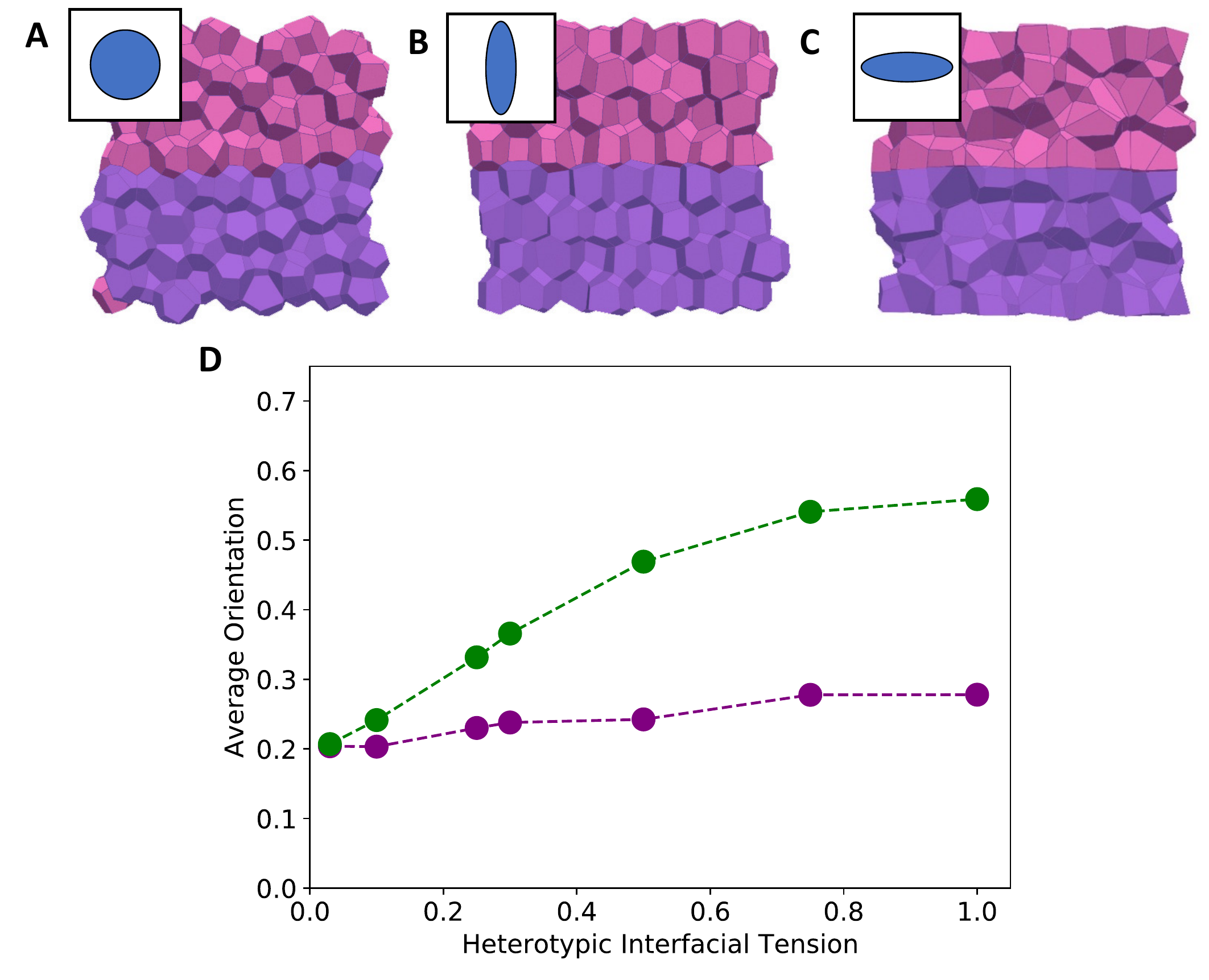}
\caption{
(A-C) Examples of tissues at different values of average orientation (0.24,0.38,0.18 for A,B,C respectively) (D) Cells will begin to orient as interfacial tension increases in the vertex model but not the Voronoi model.}
\label{fig:VertexOrientation}
\end{figure}

However, the orientation cells on the boundary exhibit a surprising difference between the two models. In the Voronoi model, as the magnitude of interfacial tension increases the cells become highly oriented perpendicular to the interface. But, in the vertex model, the cells remain randomly oriented. To quantify this orientation effect over an ensemble we define an average orientation metric.

\begin{equation}
    \langle O \rangle = \langle \frac{(\theta- \frac{\pi}{2})^2}{\frac{\pi^2}{4}} \rangle,
\end{equation}
where $\theta$ is the angle the long axis of a cell's moment of inertia tensor makes with the heterotypic interface. This metric is designed such that the alignment of all cells completely parallel to the interface yields an average orientation of zero and alignment completely perpendicular yields a value of unity. The observed range of values is slightly smaller as seen in the examples shown in Fig~\ref{fig:VertexOrientation}. Strikingly, this metric displays how sharp the difference is between the two models, where the average orientation increases steadily for the Voronoi model as HIT increases there is a negligible change in the orientation of vertex model cells. 

\subsection{Dynamic differences between models at the boundary}
So, what is causing this dramatic geometric/structural difference? For orientation to occur on the tissue boundary in Voronoi models, two things must occur; cells must remain on the interface and elongate perpendicular to the interface. In completely confluent simulations with periodic boundary conditions, for cells to elongate perpendicular to the interface they must reduce their surface area with the interface, and new cells must fill that gap. This means there must initially be a net flow of cells moving from the bulk to the interface to allow the geometry change. We speculate that this might occur if cells are kinetically pinned at the boundary, so that it is easier for them to move to the boundary than leave. The previously discussed cusp-like restoring force at heterotypic interfaces in Voronoi models~\cite{Sussman2018b} does pin cells to the boundary. We hypothesize that the magnitude of this restoring force is lower in the vertex model, which allows more frequent rearrangements at the boundary, and prevents the orientation effect.

\begin{figure}[ht!]
\includegraphics[width = 1.0\linewidth]{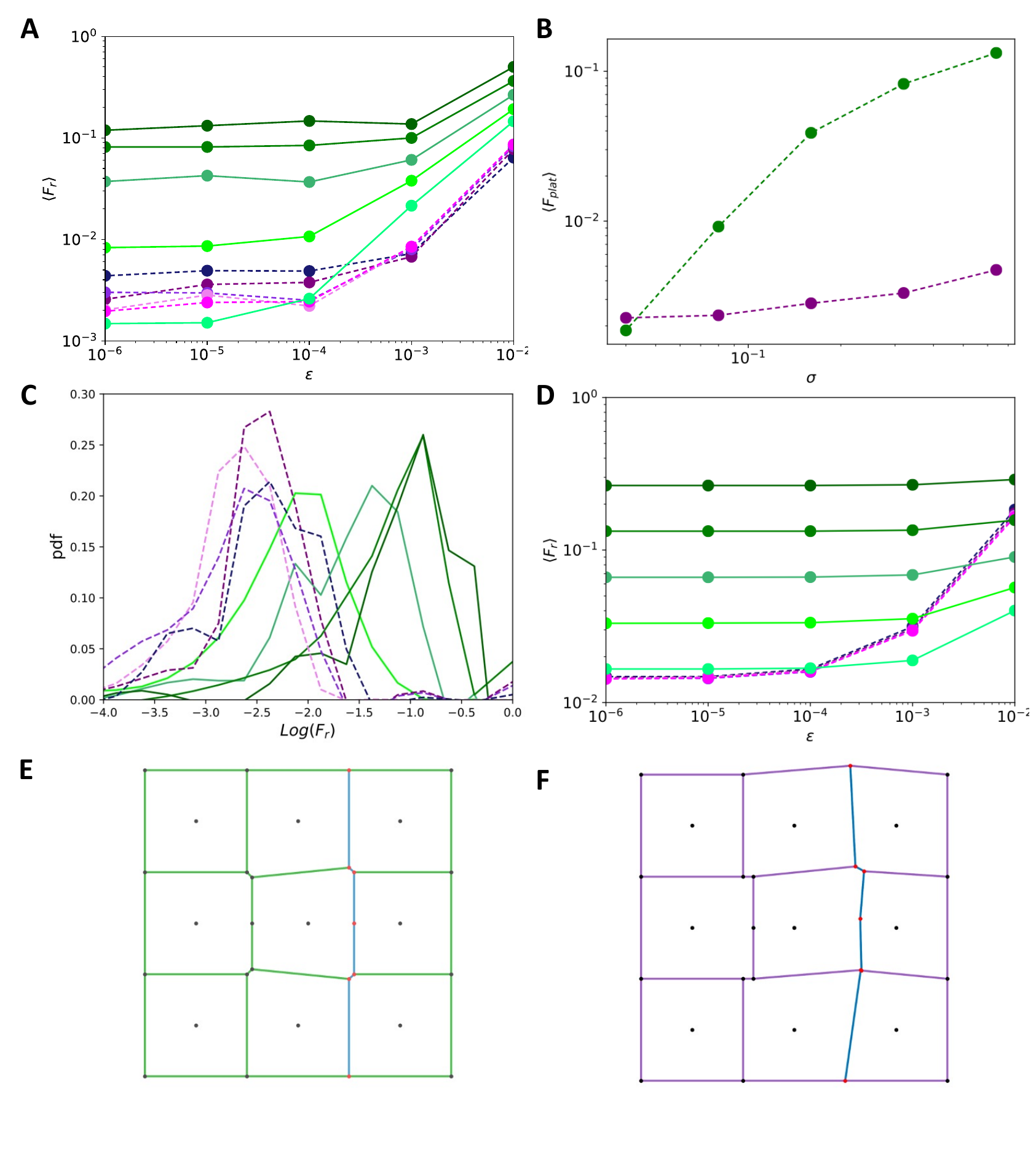}
\caption{
(A) The average cusp-like restoring force for cells, $\langle F_r \rangle$, being perturbed a distance $\epsilon$ into the heterotypic boundary. The solid green lines are from 3D Voronoi model simulations and the dashed purple lines are from the 3D vertex model where darker lines are higher HIT ($\sigma = 0.04, 0.08, 0.16, 0.32, 0.64$. (B) The plateau values extracted (A). (C) The distribution of restoring forces over an ensemble of $N=100$ systems. In both systems, the forces are normally distributed around a central peak. This data is from $\epsilon = 10^{-4}$. (D) Numeric simulations of the restoring force generated by perturbing a single Voronoi cell in 9 cell configuration in 2 dimensions (E) and perturbing the vertices of the center cell after adding random noise (This snapshot the perturbation is of magnitude $10^{-1}$ to be easily viewed) to all vertices on the interface (F). The green lines are from perturbing the Voronoi cell and the magenta lines are from perturbing the vertex model-like cell with darker lines representing higher HIT.}
\label{fig:VerVorRestoring}
\end{figure}

To test this hypothesis and measure these restoring forces, we initialize a completely segregated system of two different cell types. This system is allowed to relax over $10^5$ time steps with thermal fluctuations and then over an additional $10^6$ steps with no fluctuations to reach an energetic equilibrium. Then a single cell is selected and perturbed into the boundary with magnitude $\epsilon$. In the Voronoi model, we define the restoring force as the force on the cell center and in the vertex model as the average force on each vertex. As in previous work, we expect that a cusp in the energy will result in a restoring force that scales linearly with the interfacial tension and is independent of the displacement up to a length scale at which the normal hookean response starts to dominate. Fig~\ref{fig:VerVorRestoring}A shows precisely this response for both Voronoi (green) and vertex (magenta) models -- a flat, non-zero plateau over a range of small displacements -- demonstrating the both models exhibit a discontinuous restoring force. However, the magnitude of this restoring force increases significantly faster as function of the magnitude of the interfacial tension in the Voronoi model, as seen in Fig~\ref{fig:VerVorRestoring}B, and the plateau value in the vertex model is much less sensitive to heterotypic tension magnitude. This results in an order of magnitude difference in restoring force for moderate values of interfacial tension on the order of 0.1 times the average homotypic tension.

For a fixed displacement value, the distribution of forces over an ensemble of $N=100$ simulations in Fig~\ref{fig:VerVorRestoring}C shows an approximately normal distribution for both models. This suggests that the variation in average restoring force between the two models is not from outlier behavior in either model but a systematic difference in boundary behavior. We hypothesize that the difference in restoring force is due to the extra degrees of freedom in the vertex model that allow fluctuations at the interface to overcome the energetic barriers that protect fourfold coordinated vertices at the interface.

To make this hypothesis more concrete, we construct a simple 9-cell numerical toy model in 2 dimensions. First, we perturb a single Voronoi cell a displacement $\epsilon$ into a heterotypic interface, Fig~\ref{fig:VerVorRestoring}E, and calculate the resulting force replicating the work done in Ref~\cite{Sussman2018a}. Then to capture the variability in accessible states in the vertex model we look at the same initial 9-cell configuration but now randomly perturb the vertices on the interface with a magnitude $10^{-3}$, as seen in Fig~\ref{fig:VerVorRestoring}F. Then all the vertices of the center cell are perturbed into the interface at a displacement value $\epsilon$ and the resulting restoring force is recorded. We average the restoring force over an ensemble of $N=1000$ trials. Comparing the 2D toy model in Fig~\ref{fig:VerVorRestoring}D to the full 3D simulations in Fig~\ref{fig:VerVorRestoring}A reveals striking similarities, suggesting the 2D toy system is capturing the important features.  

\begin{figure}[ht!]
\includegraphics[width = 1.0\linewidth]{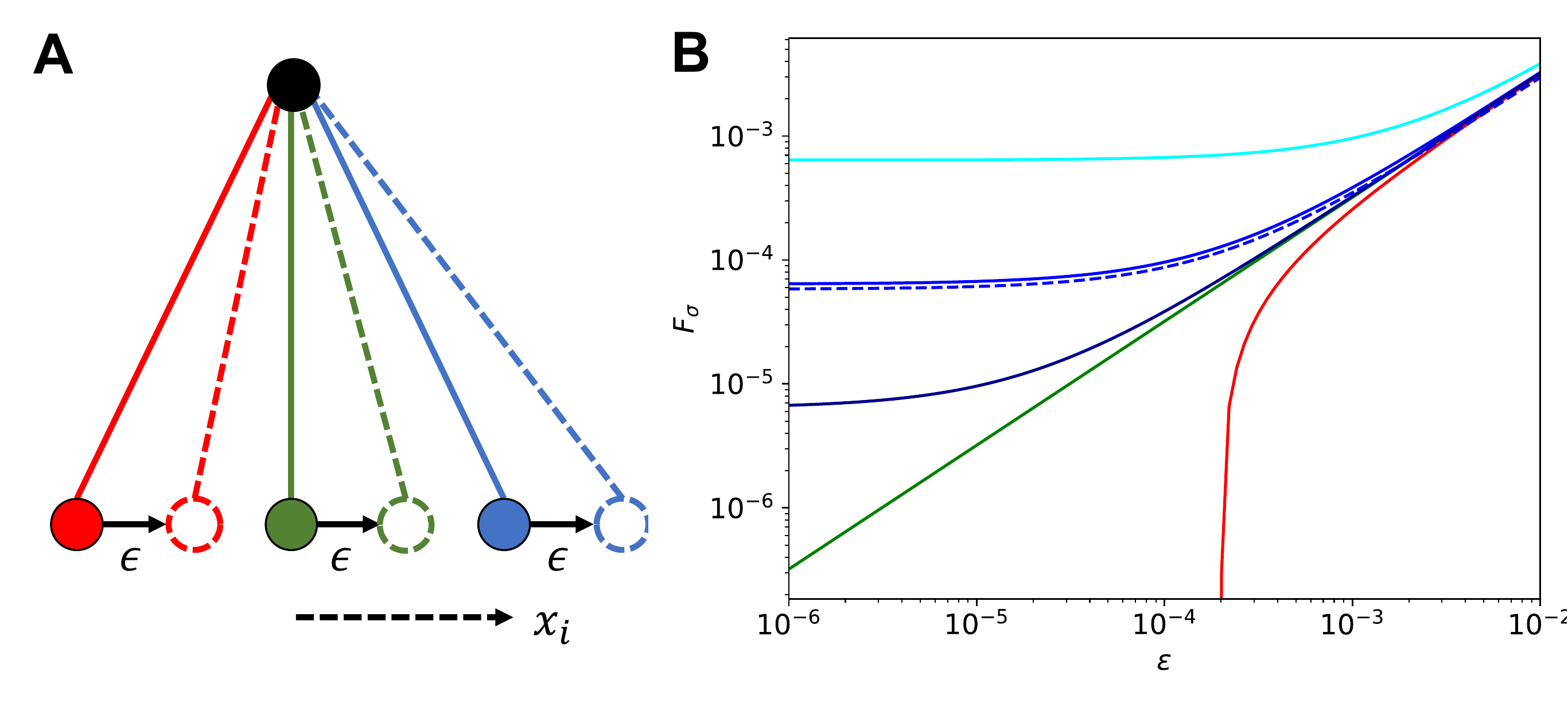}
\caption{
(A) Schematic of perturbations to single verticies on the boundary. The green line is with no perturbations, the blue is a positive perpendicular perturbation into the interface and th red is a negative perturbation. (B) Analytic restoring force due to the interfacial tension alone due to the perturbations of verticies on the interface. The solid lines represent intial perturbations of the moving cell's vertex a distance $x_i = \{-10^{-4}, 0, 10^{-5}, 10^{-4},10^{-3}\}$ colored $\{$red, green, dark blue, blue, cyan$\}$ respectively with $y_i = 1$. The dashed blue line represents a perturbation parallel to the interface such that $\{x_i,y_i\} = \{10^{-4}, 1.1\}$}
\label{fig:VertexAnalyticEdge}
\end{figure}

In the toy 2D vertex model, we can directly show that the insensitivity to HIT $\sigma$ arises because the area and perimeter terms dominate the HIT contributions to the restoring force over a wide range of HIT values $\sigma \leq 15$. This is because in vertex models, unlike Voronoi models, it is no longer the breaking of a four-fold vertex into two three-fold vertices that generates the cusp.  In fact, perfect four-fold coordinated vertices have no cusp when perturbed, as shown by the green line in Fig.~\ref{fig:VertexAnalyticEdge}. Instead, the cusps are generated by geometric nonlinearities introduced when a \emph{nearly} four-fold coordinated vertex at the end of a long edge is displaced nearly perpendicularly to that edge, as highlighted by the red and blue lines in the schematic in  Fig.~\ref{fig:VertexAnalyticEdge} (A).

Specifically, if we use coordinates where the stationary boundary point is at $\{ 0,0 \}$ and the perturbed vertex is initially at $\{x_i,y_i\}$ before being perturbed a distance $\epsilon$ into the interface we can analytically calculate the restoring force to be

\begin{equation}
    F_{\sigma} = \frac{\sigma}{\epsilon} (\sqrt{((x_i+\epsilon)^2+y_i^2}-\sqrt{x_i^2+y_i^2}).
\end{equation}

This equation confirms that if the component of the perturbation perpendicular to the interface, $x_i$, goes to zero the cusp disappears, and the restoring force scales linearly with the displacement into the interface as shown in Fig~\ref{fig:VertexAnalyticEdge}. The restoring force is negative if the vertex is perturbed away from the interface or positive if it is perturbed into the interface. Then the magnitude of the restoring force and the size of the plateau are proportional to the magnitude of the perpendicular movement. The magnitude of the restoring force in the positive direction is larger, which means for a random perturbation on average there will be a positive discontinuous restoring force proportional to the perturbation. Parallel perturbations have a negligible effect on the restoring force. This discontinuous cusp emerges due to the non-linear nature of hypotenuses.

\section{Conclusions and future work}
We have investigated differences in structure and dynamics for heterotypic interfaces between two tissue types in the 3D vertex and Voronoi models. Both models share significant similarities in demixing behavior and cell shape on the boundary and the bulk. However, registration of cell centers is significantly less in vertex models compared to Voronoi models, and cells on the boundary between tissue types will orient perpendicular to the interface in the 3D Voronoi model but not the vertex model.

In the Voronoi model, the restoring force for perturbations to cells along the boundary is an order of magnitude higher than that of the vertex model for moderate values of interfacial tension. The difference in restoring force arises from the different mechanisms that drive cusps on the boundary; in Voronoi models four-fold vertices must split into pairs of three-fold vertices in response to a perturbation, while in vertex models the cusps are created by subtle geometric nonlinearities and only arise when averaging over fluctuations, resulting in pinning forces that are much weaker. In practice, this difference means that in Voronoi models more cells can be pinned at a heterotypic interface, leading to an orientation effect not seen in vertex models.  This indicates that cell shapes at heterotypic boundaries of Voronoi models are a consequence of the representation of the degrees of freedom and not of the underlying energy functional.

As Voronoi models are significantly less computationally intensive and require fewer parameters than vertex models, this suggests that researchers should consider the dynamics and structures they are trying to resolve when choosing how to represent the degrees of freedom in a model. In simulations where dynamics near cell boundaries are not expected to play an important role, both Voronoi and vertex models generate similar mechanical and structural properties. In models that need access to more diverse cell shapes or where researchers are interested in making predictions about cell dynamics near boundaries, the 3D vertex model may be preferable.

\section{Competing interests}
The authors have declared that no competing interests exist.

\section{Acknowledgements}
The authors acknowledge financial support from the Simons Foundation grant $\#446222$ and NIH R01HD099031 (ELK and MLM) and Simons Foundation $\#454947$ (MLM), as well as a graduate dissertation fellowship from the Graduate School at Syracuse University (ELK).

\bibliographystyle{unsrt}
\bibliography{3DVertVor.bib}

\end{document}